%% file: meta_sfr.tex
\documentclass[preprint]{aastex}

\newcommand{\studynumber}{ten}
\newcommand{\fiducial}{2.7\pm 0.7}
\newcommand{\weightedmean}{2.74\pm 0.28}
\newcommand{\EdSresult}{3.3\pm 0.8}

\newcommand{\Halpha}{H$\mathrm{\alpha}$}
\newcommand{\OII}{$\mathrm{[O\,II]}$}

\begin{document}
\sloppy


\title{        A meta-analysis of cosmic star-formation history}
\author{       David W. Hogg}
\affil{        Center for Cosmology and Particle Physics,
               Department of Physics,\\
               New York University, 4 Washington Place, New York, NY 10003\\
       \textsf{david.hogg@nyu.edu}}
\slugcomment{  submitted to \textit{Pubs.~Astron.~Soc.~Pac.} 2002 August 31}

\begin{abstract}
A meta-analysis is performed of the literature on evolution in cosmic
star-formation rate density from redshift unity to the present day.
The measurements are extremely diverse, including radio, infrared, and
ultraviolet broad-band photometric indicators, and visible and
near-ultraviolet line-emission indicators.  Although there is large
scatter among indicators at any given redshift, virtually all studies
find a significant decrease from redshift unity to the present day.
This is the most heterogeneously confirmed result in the study of
galaxy evolution.  When comoving star-formation rate density is
treated as being proportional to $(1+z)^{\beta}$, the meta-analysis
gives a best-fit exponent and conservative confidence interval of
$\beta= \fiducial$ in a world model with
$(\Omega_M,\Omega_{\Lambda})=(0.3,0.7)$ and $\beta= \EdSresult$ in
$(\Omega_M,\Omega_{\Lambda})=(1.0,0.0)$.  In either case these
evolutionary trends are strong enough that the bulk of the stellar
mass at the present day ought to be in old ($>6~\mathrm{Gyr}$)
populations.
\end{abstract}


\section{Introduction}

The study of galaxy evolution is filled with negative results.
Despite intensive efforts, only small or subtle evolution has been
detected in the number density of field galaxies
\citep[eg,][]{lilly95a,heyl97,hogg98thesis,lin99a,cohen02a}, in their masses
\citep[eg,][]{vogt96,vogt97,treu99,brinchmann00,cohen02a}, or in their
clustering relative to ``stable clustering''
\citep[eg,][]{lefevre96a,carlberg97a,small99,hogg00b,carlberg00clust},
from redshift unity to the present day.  At the same time, the
constraints on galaxy evolution have not been made strong enough to
allow definitive results from the classical cosmological tests.  Has
the galaxy evolution community got anything positive to say?  Indeed
it has.

For a long time it has been observed that apparently faint and (at one
time presumed to be, now largely known to be) distant galaxies are, on
average, bluer in color than their local counterparts \citep[eg,][and
references therein]{koo92}.  This was attributed to higher
star-formation rates or younger ages at earlier times.  As the data on
distant galaxies has improved, this conclusion has been bolstered,
with photometric and spectroscopic observations spanning the full
electromagnetic spectrum.

Often, the literature on galaxy evolution focuses on disagreements
between star-formation rate measurements, as it should, since such
disagreements ought to point to important issues in selection effects,
dust extinction, the initial mass function of stars, and stellar
population syntheses
\citep[eg,][]{cowie99,bell01,hopkins01,sullivan01a}.  What is much more
remarkable than the disagreements between the measurements, however,
is the very important respect in which virtually all studies of galaxy
star-formation rates agree: The star-formation rate density in
normal galaxies has been declining from redshift unity towards the
present day.

It is of great importance that the star-formation measurements span a
number of different observational techniques, a number of different
regions in the electromagnetic spectrum, and a number of physically
different star-formation indicators.  The different studies suffer
from different selection effects and have different sensitivities to
dust extinction and the stellar initial mass function.  For example,
as the absorption due to dust in a star-forming galaxy increases, the
ultraviolet luminosity decreases, the optical \Halpha\ line decreases
by less, the radio emission from supernovae is unaffected, and the
far-infrared emission actually increases.  The different measurements
are also performed by different groups of investigators, with
different approaches and different preconceptions.  This heterogeneity
of the confirming evidence makes the decline in star-formation rate
density the most secure result in galaxy evolution, and one of the
most secure in all of astrophysics.

In what follows, a quantitative meta-analysis of the star-formation
history literature is performed, in which the declines inferred from
different observational studies are compared in a fair way, to
establish the consistency of the literature and combine the
measurements responsibly.  Because of the ``responsibility''
constraint, many studies relevant to questions of star-formation rate
have been dropped from the meta-analysis; even though almost all of
them support the final conclusions.

The star-formation rate measurements in question are measurements of
mean comoving cosmic star formation rate density $\dot{\rho}_{\ast}$,
the average mass in stars formed per unit time per unit comoving
volume.  As discussed below, except where specifically noted, all
reported measurements are corrected to a cosmological world model with
$(h,\Omega_M,\Omega_{\Lambda})=(0.7,0.3,0.7)$, where $H_0=
100\,h~\mathrm{km\,s^{-1}\,Mpc^{-1}}$ and $\Omega_M$ and
$\Omega_{\Lambda}$ are the present-day, scaled densities of matter and
cosmological constant \citep[eg,][]{hogg99cosm}.

\section{Star formation indicators}

As emphasized above, confidence in the result that star formation is
evolving comes from from the heterogeneity of star formation
indicators that show similar evolution.

\subsection{Rest-frame ultraviolet continuum}

Short-lived stars tend to have high temperatures, so the ultraviolet
continuum of galaxies is an indicator of recent star formation.  With
optical telescopes, near-ultraviolet wavelengths $\lambda\sim
2500$~\AA\ can be observed in the $U$ (or similar) bandpass at
redshifts $z> 0.4$.  A stellar population created in a short burst is
bright at $2500$~\AA\ for $\sim 20~\mathrm{Myr}$
\citep[eg,][]{bruzual93a,leitherer95} because only short-lived stars
emit at these wavelengths.

The near-ultraviolet luminosity of a galaxy is a direct measure of its
young stellar population, and therefore a good indicator of recent
star formation activity.  The primary disadvantage of this indicator
is that it is strongly affected by dust extinction; correction for
extinction requires good knowledge of the reddening law in the
near-ultraviolet and the geometry of the dust relative to the stars
\citep[eg,][]{calzetti94a}.  Another disadvantage is that ultraviolet
continuum can be produced by nuclear activity; in fact possible
contamination by nuclear activity is a disadvantage of all of the
indicators.

At redshifts $z<0.4$, the $2500$~\AA\ emission can only be estimated
from the ground by extrapolation from the visual and near-ultraviolet.
This is a disadvantage, because the extrapolation depends on the shape
of the galaxy spectral energy distribution in the ultraviolet; this is
very sensitive to small amounts of dust or small changes in star
formation history and activity.  The forthcoming GALEX mission
\citep{martin97a}, which will obtain ultraviolet imaging of normal
nearby galaxies, will solve this problem.

\subsection{\Halpha\ emission line}

The \Halpha\ line is emitted in the recombination spectrum of
hydrogen; it is a star formation indicator because only short-lived
stars produce significant luminosity in photons with energies above
the hydrogen ionization energy $13.6~\mathrm{eV}$.

Because it involves ionization and recombination, the \Halpha\ line is
in some sense a less-direct indicator of recent star formation.
However, it is weighted towards even higher-mass (and therefore
shorter-lived) stars.  It is also less affected by dust extinction
(though not completely insensitive).  The indicator has the
disadvantage that in a significant fraction of galaxies, the total
\Halpha\ luminosity can be dominated by an active nucleus.

At redshifts near unity, \Halpha\ can only be measured with
near-infrared spectrographs.  Some such instruments are now on-line,
but measurements of line strengths in faint galaxies have not been
made in large numbers.  This is likely to change in the near future.

\subsection{Near-ultraviolet emission lines}

Ultraviolet radiation from stars ionizes many atomic species, not just
hydrogen, and several of these species emit bright lines in the
optical and near-ultraviolet in their recombination spectra.  The most
prominent and useful line is the forbidden \OII\ line at 3727~\AA,
because it is bright and visible in optical spectrographs in a wide
redshift range $0.2<z<1.3$.  Because the line is forbidden, it is a
collisionally excited transition emitted by Oxygen in ionization
regions with electron densities $n\sim 10^4~\mathrm{cm^{-3}}$.  These
conditions are common around very young and forming stars.

In the local Universe, \OII\ emission appears to be closely related to
\Halpha\ emission \citep{kennicutt92a}, so it is a reasonable tracer
of star formation.  However, its strength depends on the metallicity,
density, and dust content of the nebular gas, so the relationship
between line luminosity and star formation rate is expected to have a
much greater variance than that for the \Halpha\ line.  Its primary
advantage is in redshift coverage.

The \OII\ line shares the disadvantage that it can be contaminated by
nuclear activity.

\subsection{Far-infrared continuum}

Star forming regions are often very dusty; thick dust surrounding
young stars will absorb ultraviolet and optical emission and re-emit
it at far-infrared wavelengths $30<\lambda<200~\mathrm{\mu m}$.  Since
short-lived stars are far more luminous than long-lived stars (by a
factor which overwhelms their relative scarcity in typical initial
mass functions), the far-infrared luminosity of a galaxy may be
approximately the bolometric luminosity of its dust-enshrouded,
short-lived stars.  This makes it a star-formation indicator.

Far-infrared emission is a less direct indicator than ultraviolet
flux, but it has opposite sensitivity to dust; dustier galaxies are
better measured, while pristine galaxies are missed.  It has the
enormous disadvantage that the far infrared cannot be measured
directly at ground-based telescopes.  It must be inferred from
sub-millimeter and near-infrared measurements.  The study of
far-infrared emission will be revolutionized with the SIRTF mission.
There is an additional problem that the far-infrared emission can have
a substantial contribution from an active nucleus, and the visual
extinction makes it difficult to differentiate sources of luminosity
\citep[eg][and references therein]{sanders96a}.

\subsection{Radio continuum}

A short time $\sim30~\mathrm{Myr}$ after a burst of star formation,
stars with masses $M> 8\,M_\odot$ complete most of their nuclear
burning and explode in what are expected to be Type~II and Type~Ib
supernovae, although the evidence for this association is not
iron-clad \citep[eg,][and references therein]{filippenko97a}.  These
supernovae and their remnants are bright at radio frequencies
$1<\nu<100~\mathrm{GHz}$ for $t\sim 100~\mathrm{Myr}$ \citep[eg,][and
references therein]{condon92a}.

Radio observations have the great advantage that they are insensitive
to dust.  Star formation measurement is not rocket science, but the
radio indicator is the least well-calibrated, simply because the
energetics of radio supernovae are less well-understood than other
aspects of stellar populations.  Like the other indicators, radio
measurements can be contaminated by active nuclei, which, at faint
levels, are not easily distinguished from supernovae.

\subsection{Other methods}

Many other methods have been suggested but wait for new data.  It has
been suggested that the X-ray sources created by stellar evolution
could be used as an indicator \citep{ghosh01a}.  Much work is going
into understanding the winds from forming and young stellar
populations \citep[eg,][and references therein]{kudritzki00a}; one can
imagine measuring star formation rates using mechanical, rather than
electromagnetic, luminosity.  When the SIRTF mission is able to
calibrate the relationships between strong mid-infrared spectral
features from interstellar molecules and the ultraviolet radiation
fields that excite the transitions \citep[eg,][]{li02a}, there will be
a new industry of understanding star formation activity via the
mid-infrared.

\section{Method and results}

\subsection{Studies excluded}

This meta-analysis is of star-formation rate variation \emph{within} a
single observational study, at redshifts $z<1$.  For this reason, no
study was included in this meta-analysis if it did not report more
than one independent $z\leq 1$ star-formation rate measurement.  This
excluded several otherwise relevant studies
\citep[eg,][]{connolly97,madau98a,treyer98,yan99,sullivan00,thompson01}.
An exception was made for a group of very similar surveys for
\Halpha\ line luminosity density, which were put together to make a
``combined \Halpha'' study, described below.

Several relevant studies were dropped because the targets were
selected or the results were presented such that the results could not
be fairly used as cosmic star-formation rate measurements
\citep[eg,][]{schade96,cowie97,guzman97a,lilly98a,blain99}, or because
estimates of measurement uncertainties were not provided
\citep[eg,][]{cram98}.

Only publications appearing in the refereed literature prior to 2002
August 1 were considered.

\subsection{Studies included}

The cuts left \studynumber\ studies
\citep{lilly96,hammer97,rowan-robinson97,hogg98o2lf,cowie99,flores99,mobasher99,haarsma00,jones01}.
Notes on individual studies are given in Table~\ref{tab:sfr}, but in
most cases, the confidence interval on each
star-formation-rate-density $\dot{\rho}_{\ast}$ point was measured
(with a ruler) from the relevant Figure in each paper.  The measured
points are shown in Figure~\ref{fig:sfr}.  At any redshift, there is a
large scatter.

In addition to the primary \studynumber\ studies, three very similar
studies of the cosmic \Halpha\ line luminosity density
\citep{gallego95,tresse98,glazebrook99}, each measuring the density at
a different redshift, were combined into a single ``combined \Halpha''
study, which was included as if it were an additional individual
study.

\subsection{Data fitting}

Each data point, centered on some redshift $z$, was crudely
``corrected'' to the default world model used in this paper by the
correction factor
\begin{equation}
f = \left(\frac{D_L^2}{\mathrm{d}V_C/\mathrm{d}z}\right)^{-1}\,
\left(\frac{\widetilde{D}_L^2}{\mathrm{d}\widetilde{V}_C/\mathrm{d}z}\right)
\end{equation}
where $D_L$ and $V_C$ are the luminosity distance and comoving volume
\citep[eg][]{hogg99cosm} out to $z$ in the cosmological world model
employed in the study in question, and $\widetilde{D}_L$ and
$\widetilde{V}_C$ are the same but for the fiducial world model
$(h,\Omega_M,\Omega_{\Lambda})=(0.7,0.3,0.7)$ used here.  This
world-model correction makes two assumptions: The first is that the
star-formation indicators have been calibrated somehow independently
of Hubble constant, in the sense that a change in the Hubble constant
changes the inferred luminosity and inferred star formation rate by
the same factor.  The second is that each reported data point can be
treated as well-localized at its redshift; for the precisions reported
in these studies this ought to be fine.

Linear fitting to the model $\dot{\rho}_{\ast}\propto (1+z)^{\beta}$
was performed in $\log(\dot{\rho}_{\ast})$ space with points weighted
according to their reported uncertainties.  By this procedure, the
confidence intervals reported in the individual studies are treated as
representing Gaussian uncertainties in $\log(\dot{\rho}_{\ast})$
space.  In detail, of course, this assumption is incorrect, but since
none of the studies is extremely precise and since most of the
reported uncertainties are not accurate (most only include the Poisson
contribution), the assumption does not significantly affect the
results.  The best-fit $\beta$ values and the linear-fitting
uncertainties provided by the covariance matrix are given in
Table~\ref{tab:sfr} and shown in Figure~\ref{fig:beta}.

\subsection{Results}

The weighted mean (using the inverse variances from the linear fits as
weights) of the best-fit $\beta$ values is $\beta= \weightedmean$, a
nominal ten-sigma result.

All but two of the $\beta$ measurements are consistent (at $\sim
2\,\sigma$) with the weighted mean.  In detail, the weighted mean
relies on the assumption that the studies have reported accurate
uncertainties; but weighted and unweighted mean values barely differ.
Of the two outliers, one \citep{hammer97} appears to be an outlier
simply because it carries such a small uncertainty; most of the
studies report only Poisson contributions to their error budgets, so
the uncertainties are better treated as lower limits.  The other
outlier \citep{mobasher99} is more problematic, but the sample is
small (only 23 galaxies in the $z\sim 0.5$ point), the sample is cut
on both radio power and optical magnitude (at a fairly bright
$R<21.5$~mag), and sources spectrally classified as having active
nuclei were excluded (roughly 25~percent of the sample).

Given the underestimated measurement uncertainties, a much more
conservative method for computing a mean with a reasonable confidence
interval is the bootstrap resampling technique.  In each of $10^4$
trials, ten studies were randomly chosen (with replacement) from the
group of ten and a weighted mean (using inverse variances from the
linear fits as weights) for $\beta$ was computed.  Note that each
trial will have some studies repeated, some missing.  From the $10^4$
weighted means, the lowest and highest 16~percent were discarded,
leaving the central 68-percent confidence region.  This region has a
central value and extent of $\beta= \fiducial$.  This is a
conservative but robust estimate of the average value for $\beta$.  As
described above, all measurements have been corrected to a world model
with $(\Omega_M,\Omega_{\Lambda})=(0.3,0.7)$.

If all results are transformed to an Einstein-de~Sitter Universe with
$(\Omega_M,\Omega_{\Lambda})=(1.0,0.0)$, the bootstrap-resampling
68-percent confidence interval is $\beta= \EdSresult$.

\section{Discussion}

\begin{figure}
\plotone{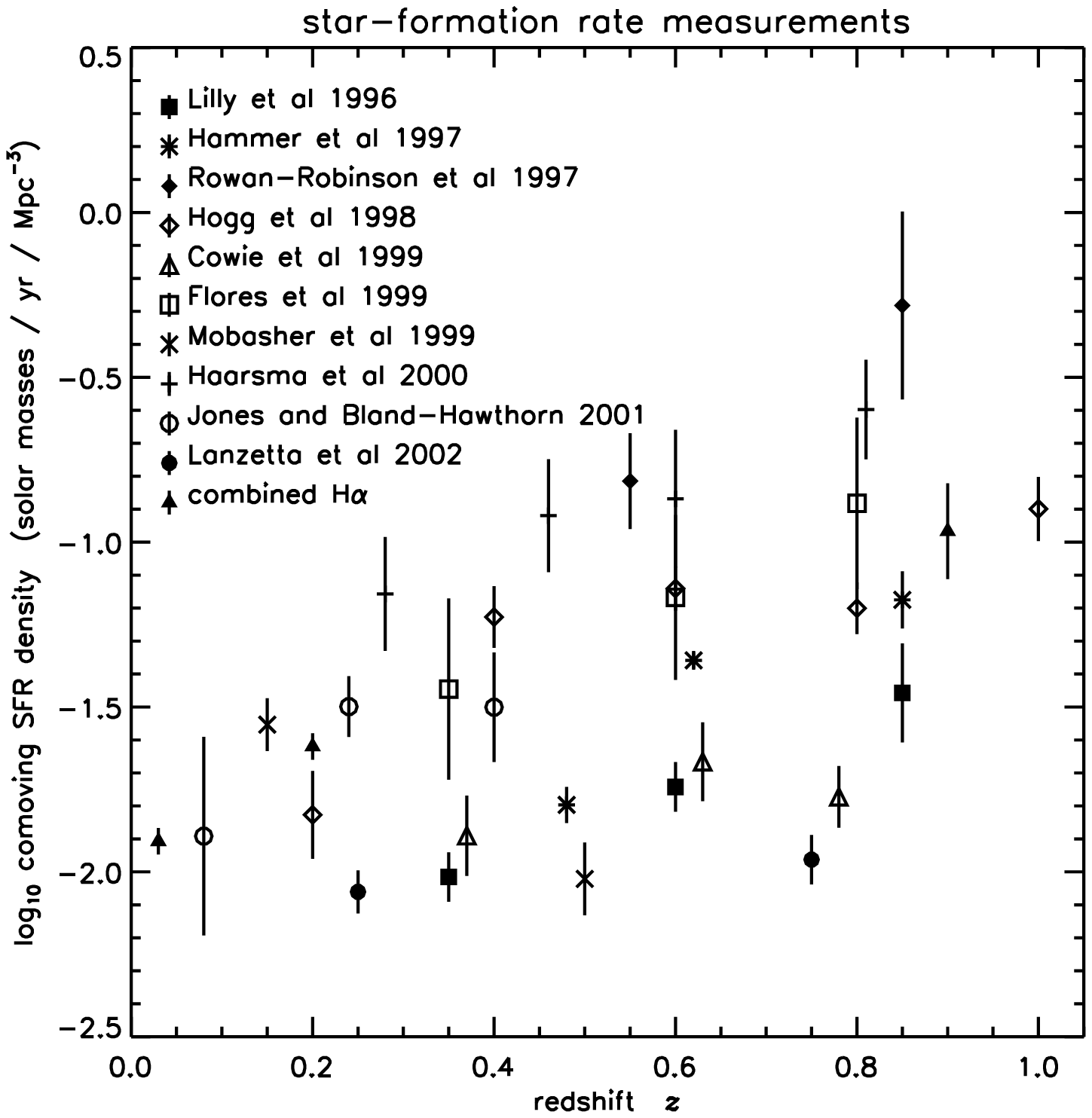}\figcaption[star formation rate measurements]{The
star-formation rate measurements used in the meta-analysis, all
corrected (crudely; see text) to a world model of
$(h,\Omega_M,\Omega_{\Lambda})=(0.7,0.3,0.7)$.  See the text and
Table~\ref{tab:sfr} for details.
\label{fig:sfr}}
\end{figure}

\begin{figure}
\plotone{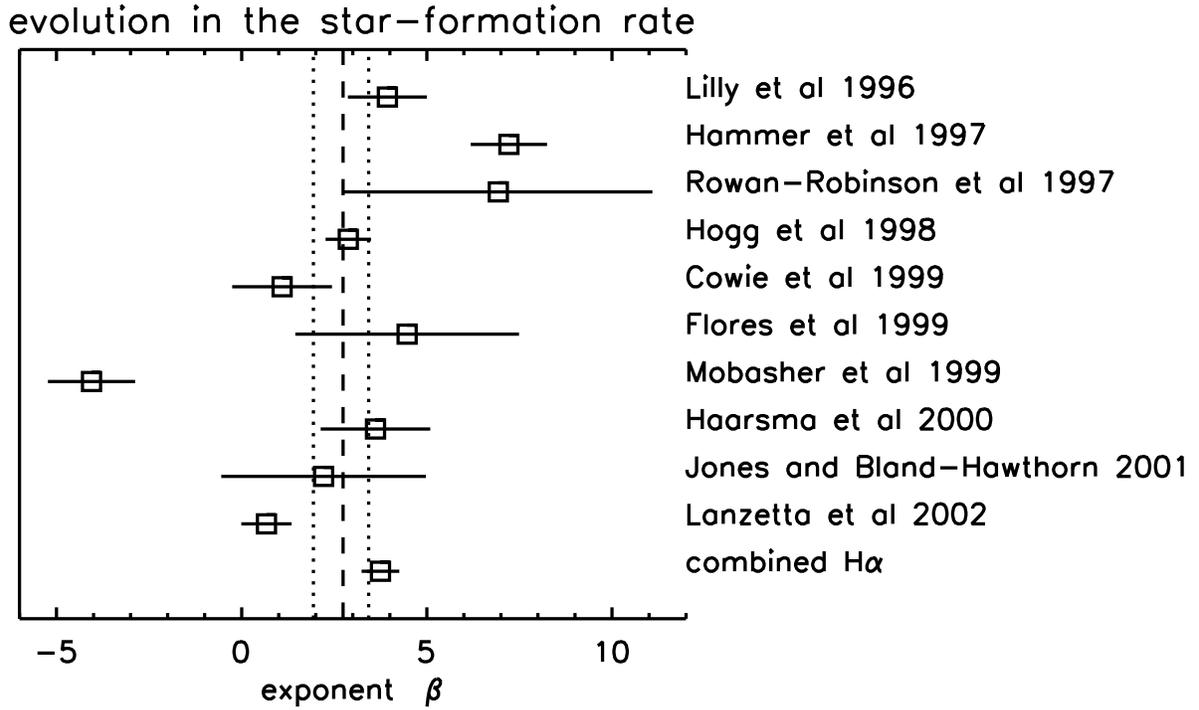}\figcaption[evolution measurements]{The best-fit
values of evolutionary exponent $\beta$ in the parameterization
$\dot{\rho}_{\ast}\propto (1+z)^\beta$ for the measurements in
Table~\ref{tab:sfr} and Figure~\ref{fig:sfr}, with world model
$(\Omega_M,\Omega_{\Lambda})=(0.3,0.7)$.  The fits are performed under
a number of assumptions described in the text.  The dashed line is the
weighted mean and the dotted lines indicate the central 68-percent
confidence interval from the bootstrap resampling described in the
text.
\label{fig:beta}}
\end{figure}

Despite the scatter in Figure~\ref{fig:sfr}, the decline in comoving
star-formation rate density $\dot{\rho}_{\ast}$ from redshift unity to
the present day is well confirmed by multiple studies.  The best
average value of the evolution exponent $\beta$, in the
parameterization $\dot{\rho}_{\ast}\propto (1+z)^{\beta}$, is $\beta=
\fiducial$.  This value is only four standard deviations from zero,
but the confidence interval is computed in the most conservative way
(resampling); the straightforward weighted mean of the measurements
gives a ten-sigma result.

In addition, almost all faint-galaxy studies which are relevant, but
for one reason or another could not be included in this meta-analysis,
bolster the conclusion that the star-formation rate has been dropping
since redshift unity.  Faint galaxies are bluer than nearby galaxies
\citep[eg,][]{koo92,smail95a,williams96}.  The surface brightnesses,
colors, and line emission of normal galaxy disks appear to have all
been dropping \citep{schade96,lilly98a}; the same appears true for blue
compact galaxies \citep{guzman97a}.  The luminosity functions of blue
galaxies and star-forming galaxies have been evolving
\citep{lilly95a,heyl97,cram98,mallen-ornelas99,lin99a}.  Models of
mid-infrared and sub-millimeter source counts may require a drop in
star-formation activity since redshift unity \citep{blain99,elbaz99}.
It is difficult to understand the observed chemical evolution of
damped Ly$\alpha$ absorbers without much higher star formation
activity at redshift unity than the present day \citep{pei95a,pei99a}.
Perhaps more indirect, the fraction of galaxies classified as
``irregular'' appears to be dropping with cosmic time
\citep[eg,][]{griffiths94,abraham96,odewahn96,vandenbergh00}, and
irregulars, locally, have above-average star-formation rates.

The mere consistency of these star-formation studies is only half the
story.  It is equally important that the studies span a wide range of
observational techniques, and a range of physically different
star-formation indicators.  This heterogeneity makes their consistency
much more impressive and much more convincing.  In particular the
different methods have different sensitivities to dust extinction and
the slope of the initial mass function of stars; it is hard to argue
that the observed trends are conspiracies of dust and mass function
variability.

\emph{The evolution in star-formation rates is the most secure result
in the field of galaxy evolution.}

Incidentally, this also makes it the strongest argument from galaxy
evolution against the pure steady-state model of cosmic origins
\citep{bondi48,hoyle48}.

If the best-fit value of $\beta$ and its conservatively estimated
uncertainty are taken at face value, then $\beta>1.3$ in the
$(\Omega_M,\Omega_{\Lambda})=(0.3,0.7)$ world model and $\beta>1.5$ in
the $(\Omega_M,\Omega_{\Lambda})=(1.0,0.0)$ world model.  In either
case, this implies that not only has the comoving star formation rate
density $\dot{\rho}_{\ast}(z)$ been declining since redshift unity,
but so has $t(z)\,\dot{\rho}_{\ast}(z)$, the comoving star formation
density per logarithmic interval in cosmic time.  This function
$t(z)\,\dot{\rho}_{\ast}(z)$ more accurately represents the formation
time of the bulk of stars.  Because $\beta$ is above the
logarithmically divergent value ($\beta >1.3$ in the default world
model), most of the stellar mass at the present day ought to be in old
($>6~\mathrm{Gyr}$) stellar populations.  This is the primary
``prediction'' of this meta-analysis; it appears to be consistent with
the data \citep{fukugita98a, hogg02a}.

In addition to observations of distant galaxies, we have another
powerful and independent ``fossil record'' of galaxy formation and
evolution: Our own Galaxy contains stellar populations formed over the
entire history of the Universe.  The star formation history of the
Galaxy can be inferred from its internal distribution of apparent
stellar ages.  Unfortunately, the distribution measured in the Solar
neighborhood is not consistent with a large decline in the Galaxy's
star formation rate over the last $\sim 8$~Gyr
\citep{barry88a,pardi94,prantzos98,rocha-pinto00,gizis02a}.  Is the
Milky Way atypical?  Or, perhaps more likely, it may be that
measurements of the age distribution in the solar neighborhood not
representative of the Galaxy as a whole; this is likely if a galaxy's
stars are formed from the inside out.  The resolution of this
inconsistency between high-redshift and Milky-Way determinations of
the star-formation rate may lead to great progress in cosmology and
galaxy evolution.

\acknowledgements It is a pleasure to thank Mike Blanton, Karl
Glazebrook, Deborah Haarsma, Heath Jones, Jim Peebles, David Schlegel,
Rosemary Wyse and Sara Zimmerman for help with the literature,
scientific discussions, and software.  This research made use of the
NASA ADS Abstract Service.

\bibliographystyle{apj}
\bibliography{apj-jour,ccpp}

\include{sfr_table}

\end{document}

%% file: sfr_table.tex
\begin{deluxetable}{lcrr}
\tablenum{1}
\tablewidth{0pc}
\tablecaption{star-formation rate measurements, corrected
\label{tab:sfr}}
\tablehead{
\colhead{\makebox[2.5in][l]{reference}} &
\colhead{} &
\colhead{} &
\colhead{}
\cr
\colhead{\makebox[2.5in][l]{technique}} &
\colhead{$z$~\tablenotemark{b}} &
\colhead{$\log_{10} \dot\rho_\ast$~\tablenotemark{c}} &
\colhead{$\beta$~\tablenotemark{d}}
\cr
\colhead{\makebox[2.5in][l]{$(h,\Omega_M,\Omega_{\Lambda})$~\tablenotemark{a}}} &
\colhead{} &
\colhead{} &
\colhead{}
}
\startdata
\cite{lilly96}~\tablenotemark{e} & $0.35$ & $-2.015 \pm  0.075$ & $ 3.93 \pm 1.07$ \\*
rest-frame ultraviolet & $0.60$ & $-1.742 \pm  0.075$ &  \\*
$(0.50,1.00,0.00)$ & $0.85$ & $-1.457 \pm  0.150$ &  \\[1ex]
\cite{hammer97} & $0.48$ & $-1.797 \pm  0.055$ & $ 7.22 \pm 1.03$ \\*
spectral features, esp.\ \OII & $0.62$ & $-1.358 \pm  0.028$ &  \\*
$(0.50,1.00,0.00)$ & $0.85$ & $-1.175 \pm  0.087$ &  \\[1ex]
\cite{rowan-robinson97} & $0.55$ & $-0.815 \pm  0.145$ & $ 6.93 \pm 4.16$ \\*
far infrared (inferred) & $0.85$ & $-0.282 \pm  0.285$ &  \\*
$(0.50,1.00,0.00)$ &  &  &  \\[1ex]
\cite{hogg98o2lf} & $0.20$ & $-1.827 \pm  0.133$ & $ 2.88 \pm 0.61$ \\*
\OII\ line emission & $0.40$ & $-1.227 \pm  0.093$ &  \\*
$(1.00,0.30,0.00)$ & $0.60$ & $-1.141 \pm  0.067$ &  \\*
 & $0.80$ & $-1.200 \pm  0.078$ &  \\*
 & $1.00$ & $-0.899 \pm  0.097$ &  \\[1ex]
\cite{cowie99}~\tablenotemark{f} & $0.37$ & $-1.890 \pm  0.122$ & $ 1.09 \pm 1.35$ \\*
rest-frame ultraviolet & $0.63$ & $-1.666 \pm  0.119$ &  \\*
$(0.65,1.00,0.00)$ & $0.78$ & $-1.773 \pm  0.094$ &  \\[1ex]
\cite{flores99} & $0.35$ & $-1.445 \pm  0.275$ & $ 4.47 \pm 3.02$ \\*
far infrared (inferred) & $0.60$ & $-1.167 \pm  0.250$ &  \\*
$(0.50,1.00,0.00)$ & $0.80$ & $-0.882 \pm  0.260$ &  \\[1ex]
\cite{mobasher99}~\tablenotemark{g} & $0.15$ & $-1.553 \pm  0.080$ & $-4.05 \pm 1.18$ \\*
1.4 GHz radio & $0.50$ & $-2.021 \pm  0.110$ &  \\*
$(0.50,1.00,0.00)$ &  &  &  \\[1ex]
\cite{haarsma00} & $0.28$ & $-1.157 \pm  0.173$ & $ 3.61 \pm 1.48$ \\*
1.4 GHz radio & $0.46$ & $-0.920 \pm  0.171$ &  \\*
$(0.50,1.00,0.00)$ & $0.60$ & $-0.869 \pm  0.209$ &  \\*
 & $0.81$ & $-0.598 \pm  0.151$ &  \\[1ex]
\cite{jones01}~\tablenotemark{h} & $0.08$ & $-1.892 \pm  0.301$ & $ 2.21 \pm 2.76$ \\*
\Halpha\ line emission & $0.24$ & $-1.499 \pm  0.092$ &  \\*
$(0.50,1.00,0.00)$ & $0.40$ & $-1.501 \pm  0.166$ &  \\[1ex]
\cite{lanzetta02}~\tablenotemark{i} & $0.25$ & $-2.061 \pm  0.065$ & $ 0.67 \pm 0.68$ \\*
rest-frame ultraviolet & $0.75$ & $-1.963 \pm  0.075$ &  \\*
$(1.00,1.00,0.00)$ &  &  &  \\[1ex]
combined \Halpha~\tablenotemark{j} & $0.03$ & $-1.907 \pm  0.040$ & $ 3.75 \pm 0.51$ \\*
\Halpha\ line emission & $0.20$ & $-1.620 \pm  0.040$ &  \\*
$(0.50,1.00,0.00)$ & $0.90$ & $-0.967 \pm  0.145$ &  \\[1ex]
\\
\multicolumn{3}{l}{weighted mean} & $ 2.74 \pm 0.28$ \\
\multicolumn{3}{l}{bootstrap resampling~\tablenotemark{k}} & $ 2.68 \pm 0.75$ \\
\enddata
\tablenotetext{a}{Cosmological parameters used in the original
reference.  For the purposes of fitting, the results have been
corrected to an effective world model of
$(h,\Omega_M,\Omega_{\Lambda})=(0.7,0.3,0.7)$;
see the text for details.}
\tablenotetext{b}{$z$ is redshift.
Only points at redshifts $z\leq 1$ were included in the fits.}
\tablenotetext{c}{$\dot\rho_\ast$ is the comoving star-formation
rate density.
All measurements have been corrected to the fiducial world model
as described in the text.}
\tablenotetext{d}{$\beta$ is the exponent in parameterization
$\dot\rho_\ast\propto (1+z)^{\beta}$.
See text for explanation of the fitting technique
and calculation of uncertainties.}
\tablenotetext{e}{Data for \cite{lilly96} are taken from
analysis of \cite{glazebrook99}.}
\tablenotetext{f}{Luminosity density in \cite{cowie99} has been
converted to star-formation rate with
the conversion of \cite{cowie97}.}
\tablenotetext{g}{\cite{mobasher99} do not give values for the
cosmological parameters; I have \emph{guessed} (0.5,1.0,0.0).}
\tablenotetext{h}{Integrations of the functions in \cite{jones01}
were kindly provided by Jones (private communication).}
\tablenotetext{i}{The ``scaled break intensity'' results of
\cite{lanzetta02} are used here.}
\tablenotetext{j}{The ``combined \Halpha'' line is a compilation
of three similar studies by \cite{gallego95}, \cite{tresse98}, and
\cite{glazebrook99}, with star-formation analysis by
\cite{glazebrook99}.}
\tablenotetext{k}{The central value and uncertainty reported for the
bootstrap resampling indicate the central 68-percent confidence
interval; see the text for details.}
\end{deluxetable}